# Polarity from the Bottom Up: A Computational Framework for Predicting Spontaneous Polar Order

Jordan Hobbs [1], Calum J. Gibb [2], Richard J. Mandle [1,2,*]

* r.mandle@leeds.ac.uk
[1] School of Physics and Astronomy, University of Leeds, Leeds, LS2 9JT
[2] School of Chemistry, University of Leeds, Leeds, LS2 9JT

## Abstract

So-called polar liquid crystals possess spontaneous long-range mutual orientation of their electric dipole moments, conferring bulk polarity to fluid phases of matter. The combination of polarity and fluidity leads to complex phase behaviour, and rich new physics, yet the limited understanding around how specific molecular features generate long-range polar ordering in a fluid is a hindrance to development of new materials. In this work, we introduce a computational framework that probes the bimolecular potential energy landscape of candidate molecules, enabling us to dissect the role of directional intermolecular interactions in establishing polar order. In closely related families of materials we find conflicting preferences for (anti)parallel ordering which can be accounted for by specific interactions between molecules. Thus, our results allow us to argue that the presence (or absence) of polar order is a product of specific molecular features and strong directional intermolecular interactions rather than being simply a product of dipole-dipole forces. The design principles established can be leveraged to developing new polar liquid crystalline materials.

## Introduction

Liquid crystals, known for almost 150 years, are characterised by orientational (and positional) ordering of their constituent molecules or particles (e.g. the nematic phase, Figure 1a). Within these, there is typically no preference for long-range orientation of the molecular electric dipole moment, and so the bulk phases are *apolar*. In recently discovered polar liquid crystals there is a strong preference for parallel orientation of the electric dipole moments of the constituents. In the simplest case, the ferroelectric nematic ($N_F$; Figure 1b), [1-4] conventional nematic orientational order is augmented by an additional polar order, which manifests as the constituent molecules having their electric dipole moments oriented parallel and so endowing the bulk phase with spontaneous polarisation. The $N_F$ phase - combine the magnitude of polarisation of solid-state materials (e.g. the spontaneous polarisation of RM734 is ~ 0.06 C m$^{-2}$ versus ~0.8 C m$^{-2}$ for $LiNbO_3$ [5]) with the fluidity and ease of processing of liquids/fluids, and so their discovery has been met with great enthusiasm for possible future applications. [6-8]

Since this initial discovery, the imposing of polar order atop liquid crystal with differing positional order has been shown to yield orthogonal (SmA$_F$) [9,10] and tilted (SmC$_P$) phases. [11,12,13] More complex phase types are also shown to exist, including spontaneously helical nematic ($^{HC}$N$_F$ in [14] or N$_{TBF}$ in [15]) and smectic (SmC$_P^H$). [16] It is reasonable to expect, where not forbidden by symmetry, polar variants of classical liquid crystals will be discovered. For example, higher ordered orthogonal (SmB$_P$, SmE$_P$) or tilted (SmI$_P$, SmF$_P$) smectics, where the subscript P denotes polar order being preferred to "F" for ferroelectric as the viscosity of such hypothetical states may render the measurement of polarization reversal ambiguous.

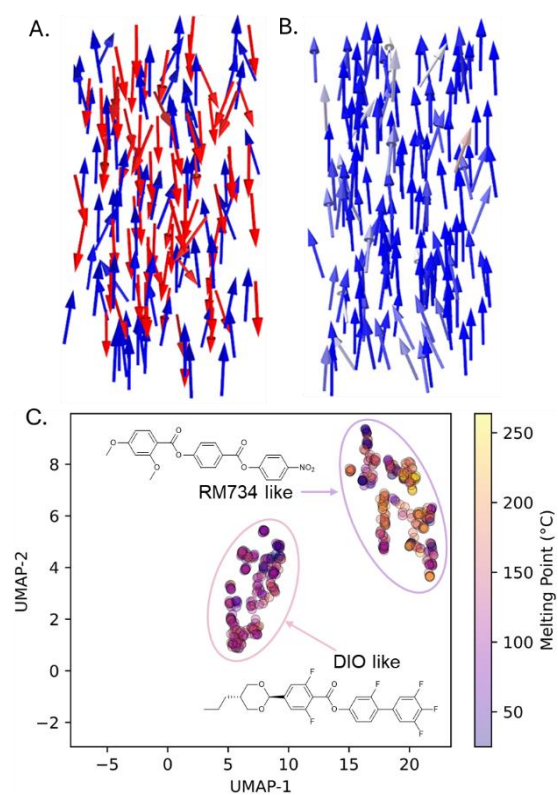

**Figure 1:** Depictions of *apolar* (A) and *polar* (B) nematic phases; the arrows represent molecular electric dipole moments and are coloured according to their orientation. (C) Molecular similarity of polar liquid crystals visualised with a plot of Uniform Manifold Approximation and Projection (UMAP) – molecules are represented as a 1024-bit Morgan fingerprint [17] with a radius of 2, 50 nearest neighbours, a minimum distance of 0.1, and the Roger-Stanimo similarity metric to project the high-dimensional fingerprint to two dimensions. To aid visualisation, each datapoint is coloured by the melting point of its parent molecule. The plot encodes data for ~ 500 polar liquid crystals, with data taken from refs. [9, 16, 18-44]

While typified by archetypal materials such as RM734 [45, 46] and DIO, the $N_F$ phase is fairly widespread and found in an ever growing number of materials. [42, 47] Despite this, there is no unified view of *how* polar order arises. The effect is that while a few materials are discovered *ad hoc,* most are progressions or evolutions from some initial configuration (Figure 1c), rather than designed from first principles. The molecules that show polar liquid crystal phases typically have large electric dipole moments, and it has been suggested that dipole-dipole interactions are important in the genesis of polar order. However, small changes in molecular structure, even those which preserve or even enhance the electrical polarity of the molecule, can lead to the loss of polar ordering. [11, 47] In our view, dipole-dipole interactions are likely important, but an argument could be made that their importance is currently overstated where more recently it has been demonstrated that strong dipole-dipole interactions can lead to anti-parallel ordering [48, 49] and twisted structures even with degenerate anchoring. [50-52]

Linking emergent polar order to *specific patterns of intermolecular interactions,* similar to the Madhusudhana charge interaction [53] but grounded in atomistic detail, is not only more logical from a molecular viewpoint, but has the potential to impact other adjacent fields with polar order (for example, polymers, MOFs and such).

Atomistic MD simulations have given valuable insight into the molecular basis of polar order in liquid crystal phases. [54] However, the reliance on empirical force fields neglects electronic effects such as polarization, charge transfer, and intermolecular induction effects. Recently, we developed an analysis workflow for examining the 3D bimolecular potential energy surface (bPES) for an interacting pair of molecules, [47] this being particularly suited to the study of polar liquid crystals. Herein, we show that the presence and absence of the $N_F$ phase in closely related materials (RM734 vs RM734-CN; Fig 2) can be accounted for by solely considering lateral intermolecular interactions. We extend this analysis workflow DIO, [22] and the closely related materials CIO [43] and DIO(-F). [42]

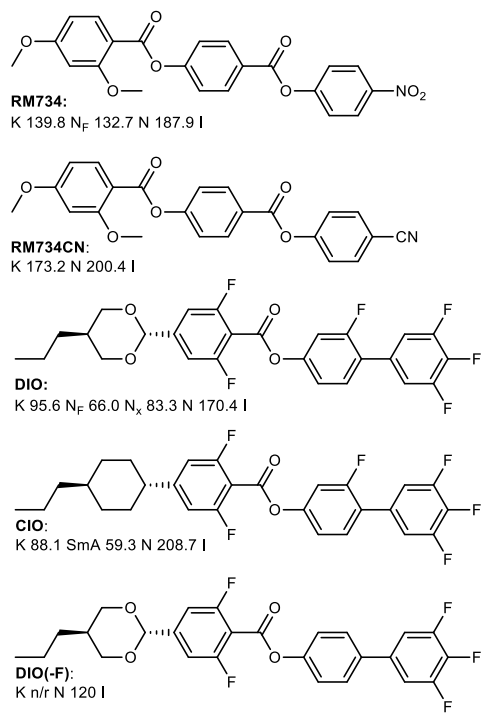

**Figure 2:** Molecular structure of the materials used in this study.

**Methods**

Transition temperatures were taken from the literature: RM734 and RM734-CN were taken from Mandle *et al*. [20, 21]; DIO from Nishikawa *et al*. [22]; CIO from Hobbs *et al*. [43]; DIO(-F) from Aya *et al*. [42] Electronic structure calculations were performed in ORCA 5.0.4. [55, 56] The rigid bimolecular potential energy surface was constructed as follows, using the software tool described in ref [47].

First, the geometry of a single molecule was optimised at the B97-D3/cc-pVTZ level of DFT. [57, 58] A series of rigid bimolecular calculations were set up by translating a second copy of this molecular geometry over the x/y/z axes – here, we restrict our study to the limiting cases of parallel and antiparallel orientation of the two molecules. The initial displacement grid has an initial spatial extent of x = ±35 Å; y = ±20 Å; z = ±8 Å, with a resolution of 1 Å.

We impose geometric constraints on the intermolecular separation by enforcing a minimum and maximum atom-atom distance criterion. Specifically, configurations are only considered valid if the shortest interatomic distance between any atom in molecule 1 and any atom in molecule 2 is no less than 3.0 Å, ensuring physically reasonable non-covalent interactions. Conversely, to exclude unbound or weakly interacting states, we impose an upper bound of 5.0 Å on the longest interatomic separation between the two molecules. These criteria ensure that the sampled configurations remain within a physically relevant interaction range while preventing artificial close contacts or excessively weak long-range interactions. For each unique geometry we perform a single point energy calculation at the B97-D3/cc-pVTZ level of DFT with counterpoise correction, with the complexation energy and displacement vectors used to construct the 3D bimolecular potential energy surface.

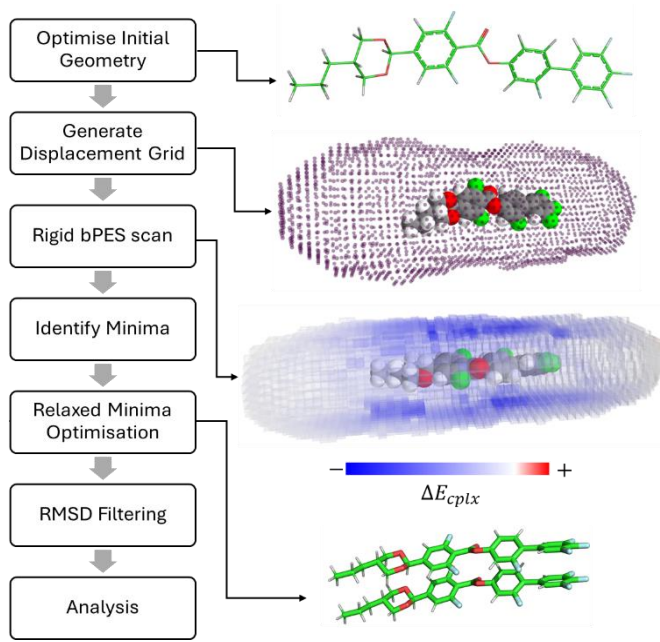

**Figure 3:** A schematic overview of the workflow involved in this analysis

We identify the global minimum on the bPES, along with a further 9 discrete local minima. Minima are only considered discrete if they are 3 Å from another identified local minima and are not related via symmetry. Each identified minima then undergoes unconstrained geometry optimisation at the same level of DFT as used previously. To account for optimisations of two (or more) configurations converging to the same geometry, we calculate the RMSD between each set of optimised minima geometries for each molecule and discard those that fall below a threshold of 0.2 Å. Visual inspection of remaining geometries is used to classify specific pairing types (e.g. slipped parallel, twisted antiparallel and so on). A schematic of this workflow is given in Figure 3.

**Results**

We begin with a general observation on the obtained rigid potential energy surfaces. The long-range orientational correlation of electric dipole moments in polar liquid crystals inevitably leads to invocation of head-to-tail interactions between molecules as a rationale. Chemically, this is dubious - the non-polar "tail" of the molecule (a saturated hydrocarbon) will not experience particularly strong interactions with the polar "head" (a nitro group, a nitrile etc.). Here, we find the energetics of head-to-tail interactions in the molecules studied here to be negligible, with $\Delta G_{complex}$ being smaller than 1 kcal / mol. As expected, head-to-head (or tail-to-tail) interactions where polar functionalities on adjacent molecules are proximal leading to electrostatic repulsion (Fig 4a). Instead, the results below demonstrate that polar order (parallel molecule arrangement) is the result of rather more subtle and directional non-covalent forces.

We first consider the well-known materials RM734 and RM734-CN. Whereas RM734 has a terminal nitro group and exhibits the $N_F$ phase, in RM734-CN the nitro group is swapped for a cyano group which leads to the absence of the $N_F$ phase. Given that the two materials have comparable electric dipole moments (11.4 D versus 11.3 D, for RM734 and RM734-CN respectively, at the B97-D3/cc-pVTZ level of DFT) the magnitude of electric dipole should not be taken exclusively as indicative of spontaneous polar ordering.

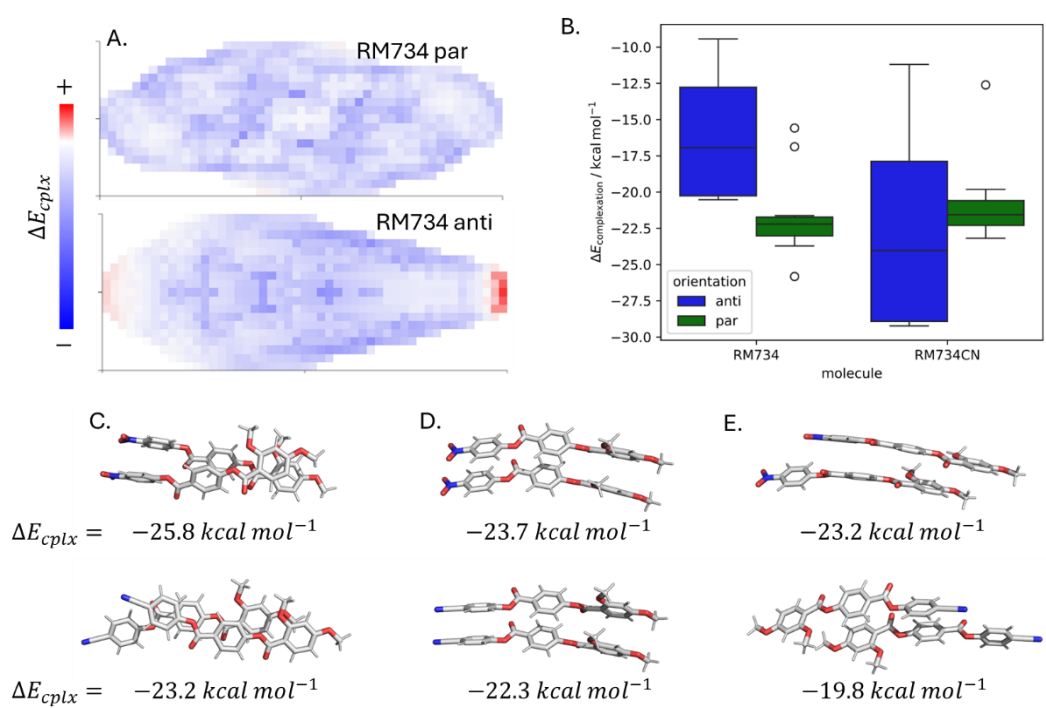

**Figure 4:** (A) The rigid bimolecular potential energy surface for RM734 in parallel (top) and antiparallel configurations; (B) boxplot of complexation energies of 10 discrete minima subjected to unconstrained geometry optimisation; Comparison of principal pairing modes of RM734 (top) and RM734CN (bottom) and the counterpoise corrected DFT:B97-D3/cc-pVTZ complexation energies: The twisted-parallel form (C); the parallel form (D); the slipped parallel form (E).

The rigid bimolecular potential energy surfaces for the two materials are not especially diagnostic upon visual examination; both the parallel and antiparallel configurations feature many local minima (Figure 4a). However, following geometry optimisation of each minimum, we find significant differences between RM734 and RM734CN. For RM734, the parallel form is the global minimum, with multiple low energy parallel configurations whereas for RM734CN the antiparallel form is the global minimum (Fig 4b). Broadly speaking, any parallel configuration of RM734 is slightly lower in energy than the equivalent configuration in RM734CN (Fig 4c-e), but the difference in complexation energy is modest (~ 1-2 kcal mol$^{-1}$). The global energy minima for RM734 is the twisted-parallel form (Fig 4c), featuring a weak electrostatic interaction between staggered nitro groups on adjacent molecules $(O_2N^{\delta^+} \cdots O^{\delta^-}NO)$ which is not feasible for the nitrile terminated material in this geometry. In the parallel form (Fig 4d), the same interaction is present for the nitro-terminated material, while for the nitrile terminated RM734-CN there is an analogous but weaker interaction $(C \equiv N^{\delta^-} \cdots C^{\delta^+} \equiv N)$, which reflects in the reduced complexation energy. In the case of the slipped parallel form (Fig 4e), the geometry of the nitro group leads to a favourable interaction with the proximal carboxylate ester on the adjacent molecule $(O = C^{\delta^+} \cdots O^{\delta^-}NO)$ which is not geometrically possible for the nitrile terminated material. Thus, the preference for parallel (polar) pairing modes results from the subtle balance of geometry with non-covalent interactions.

The behaviour of antiparallel configurations is somewhat more nuanced. On the one hand, some forms (such as the slipped antiparallel pair in Figure 5a) are close in terms of geometry and complexation energy. However, the nitrile terminated material adopts certain antiparallel configurations (such as the global energy minimum twisted antiparallel form, Fig 5d) which are not found for the nitro terminated material. The reason for this can be understood in terms of charge distribution (Fig 5e-f); both nitro- and cyano-groups are highly polar with localised charges, however the geometry of the nitro group - with large, localised charges on the two oxygen atoms - prevents the twisted antiparallel configuration due to electrostatic repulsion from the oxygen atom of the terminal methoxy unit. Thus, we are left with the counterintuitive conclusion that polar order in RM734, and absence of polar order in RM734-CN, results from a small increase in complexation energy for the parallel form and a significant decrease for the antiparallel form.

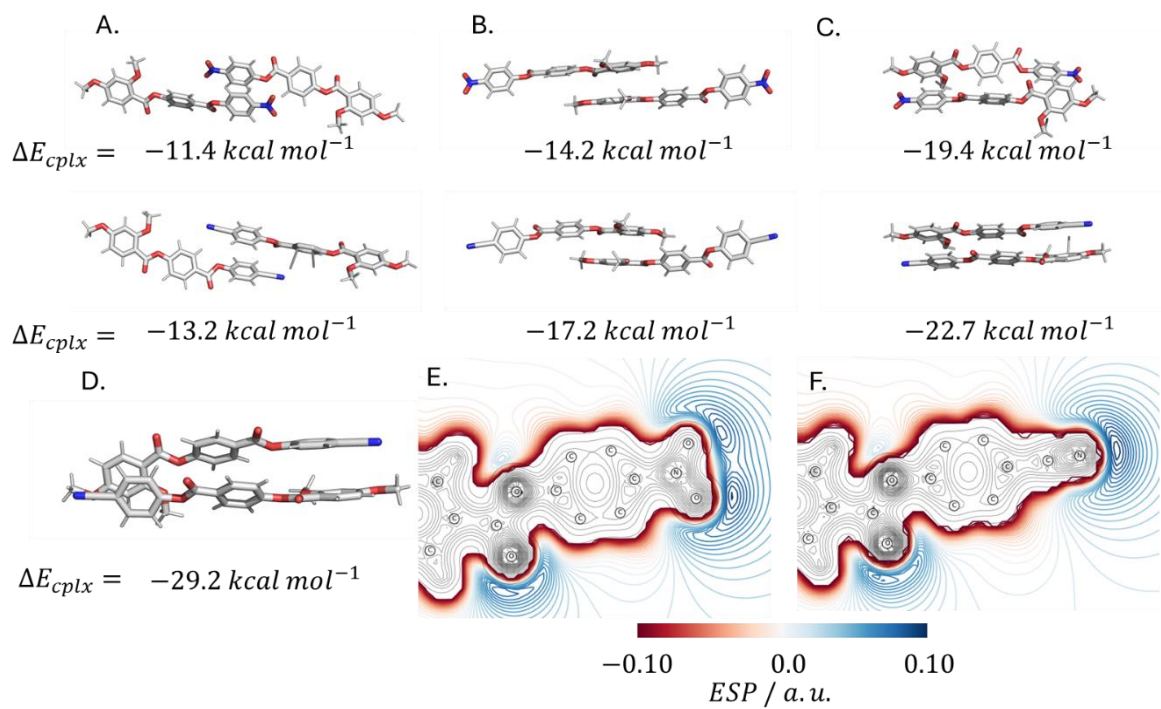

**Figure 5:** Comparison of principal antiparallel pairing modes of RM734 (top) and RM734CN (bottom) and the counterpoise corrected DFT:B97-D3/cc-pVTZ complexation energies: major slipped antiparallel pair (A); minor slipped antiparallel pair (B); antiparallel pair (C). The twisted antiparallel pair of RM734-CN (D) is the global energy minimum for this material, yet this configuration is not formed by the nitro terminated RM734. Contour plots of ESP projected onto the molecular plane for RM734 (E) and RM734CN (F); the geometry of the nitro group results in a larger charge volume than the nitrile.

We now consider DIO and two derivatives; CIO [43] and DIO(-F). [42] Replacement of the *trans* 1,3-dioxane unit of DIO with a *trans* cyclohexyl unit affords CIO, which exhibits only apolar liquid crystal phases. Similarly, DIO(-F) - analogous to DIO but with a single fluorine atom removed – displays only apolar liquid crystalline order. As with the previous examples, the overall polarity of these materials is comparable, with calculated dipole moments at the DFT:B97-D3/cc-pvTZ level of 8.5 D (DIO), 7.0 D (CIO) and 8.1 D (DIO(-F)).

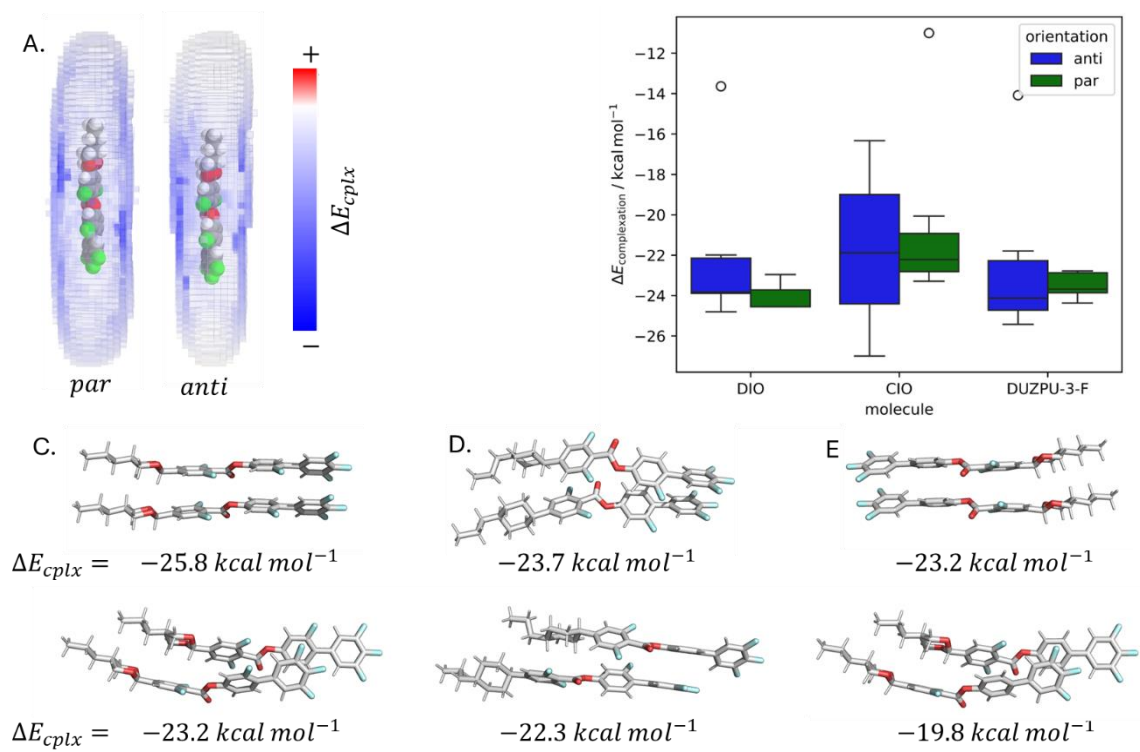

**Figure 6:** Rigid bimolecular PES for DIO in parallel and antiparallel configurations (A); boxplot of complexation energies of 10 discrete minima subjected to unconstrained geometry optimisation (B); Minimum energy configurations of DIO (C), CIO (D), and DIO(-F) (E) each in the parallel (top) and slipped parallel (bottom) configurations. $\Delta E_{cplx}$ is the counterpoise complexation energy at the DFT B97D3/cc-pVTZ level.

In the parallel configuration, all three materials adopt parallel (Fig 6c-e, top) and slipped parallel (Fig 6c-e, bottom) configurations. For both configurations, the complexation energy of DIO is larger than CIO or DIO(-F). In the first case, this reflects the contribution of the dioxane group (DIO, versus cyclohexane in CIO); with the dioxane group interacting with either another dioxane (parallel) or the adjacent *meta* difluorobenzene (slipped parallel). This interaction is via electrostatics, with the partial negative charges of the oxygen "side" of the dioxane ring being proximal to the partial positive charge of an adjacent dioxane or benzene ring. It should also be noted that dioxane has a smaller steric footprint than cyclohexane, which may also be a relevant factor. We speculate that the incapability of other saturated ring systems to deliver polar liquid crystal phase (e.g. 1,3,2-dioxaborinanes, cubanes) [43] is due to similarly reduced interaction strength (and potentially steric factors also) versus the parent dioxane. Based on this, we would expect the tetrahydropyran homologue of DIO to have a somewhat lower onset temperature for polar order. The analogous 2,6,7-trioxabicyclo[2.2.2]octyl material is also likely to be lower owing to its larger steric footprint and unfavourable geometry. The 1,3-dioxane ring, while prone to degradation (hydrolysis, isomerisation, thermolysis) [43], does regrettably appear to be a privileged scaffold with respect to polar liquid crystals.

Comparing DIO with DIO(-F), the reduction in complexation energy reflects the impact of fluorination pattern on π- π interactions, the strength and geometry of which reflects the anisotropy of the electrostatic potential (ESP) and the quadrupole moment of the system. It is well known that benzene has a negative quadrupole moment perpendicular to the ring plane owing to delocalised electrons, while hexafluorobenzene has a positive quadrupole as the fluorine atoms pull electron density away from the ring π-cloud. This effect also manifests in the ESP, with benzene having a negative ESP above/below the ring due to the density of π-electrons, and hexafluorobenzene having positive ESP above/below the ring. Thus, the heterocomplex of benzene and hexafluorobenzene can stack face-to-face due to complementary ESP. In our previous work, [16] we demonstrated that the ESP should be spatially uniform in 3D. We can use the example of a black and white checkboard pattern to illustrate this in a simplistic fashion, where black and white are alternate charge polarities. When the checkboard pattern gets disturbed (i.e. a white square is turned black) a region of repulsion is induced reducing the prevalence of lateral interactions. Induce enough of these repulsive regions and the anti-parallel promoting dipole-dipole interactions becomes dominant once again.

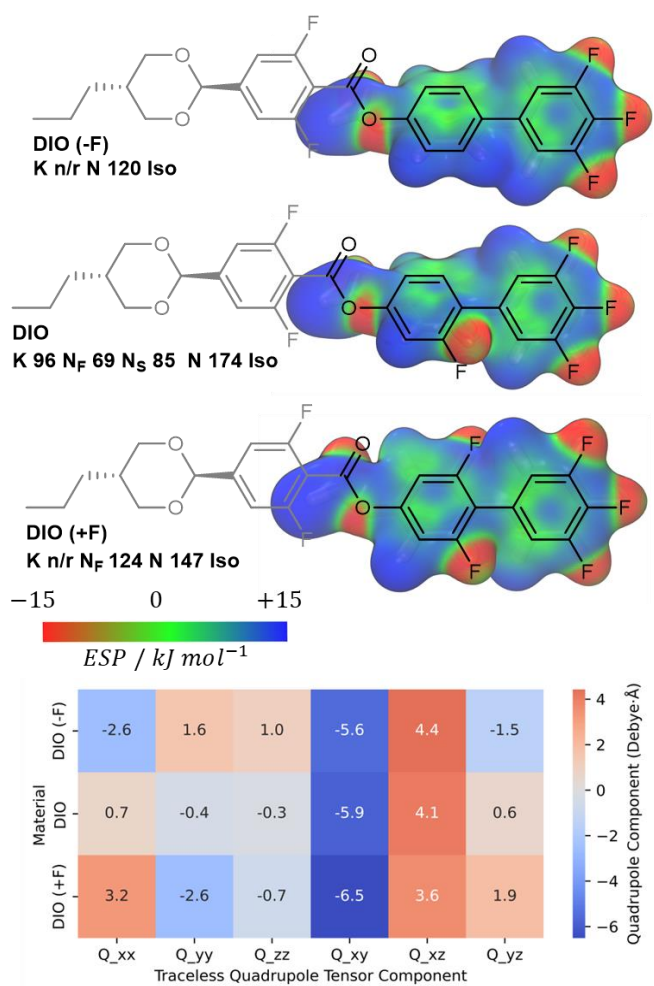

**Figure 7:** Molecular structures and transition temperatures (°C) of DIO, DIO (-F), and DIO (+F). [42] Right, calculated ESP potential mapped onto the electron density at the 0.04

isosurface calculated for the indicated portion of each molecule. Bottom, traceless quadrupole tensor components for each material, with the molecular long axis oriented along X. All calculations were performed at the DFT:B97D3/cc-pVTZ level.

Here, the fluorination pattern in DIO leads to a more anisotropic ESP within the rings than for DIO(-F), favouring face-to-face stacking through complementary ESP between adjacent molecules. The ESP anisotropy is increased by incorporating an additional fluorine atom (DIO (+F); Fig 7); this is expected to further enhance the packing preferences, and manifests experimentally as an increased $T_{(NF)}$. The effect of fluorination pattern can also be seen to impact on the quadrupole tensor, with the $Q_{YY}$ and $Q_{ZZ}$ components changing sign as the degree of fluorination increases. This in itself is not proscriptive for determining the preference for parallel/antiparallel packing; geometry and repulsion between localised charges (e.g. on fluorine) must also be considered.

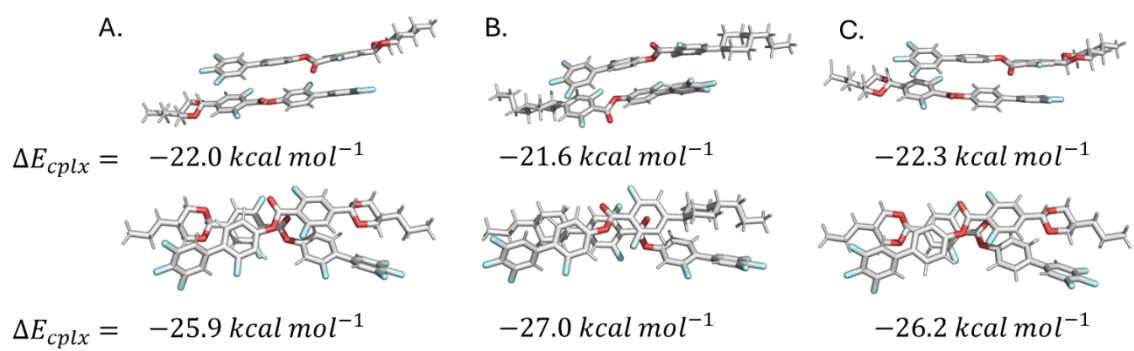

**Figure 8:** Minimum energy configurations of DIO in the slipped antiparalell (top) and twisted antiparalell (bottom) configurations (A); CIO in slipped antiparallel (top) and twisted antiparalell (bottom)configurations (B); DIO(-F) in the slipped antiparallel (top) and twisted antiparalell (bottom) (C). $\Delta E_{cplx}$ is the counterpoise complexation energy at the DFT B97D3/cc-pVTZ level.

We next consider the case of antiparallel geometries. All three materials exhibit the same types of pairing modes, with the twisted antiparallel pair (Fig 8, bottom) being the global energy minimum antiparallel form. DIO, and its derivatives considered here, show smaller differences in terms of the complexation energy of antiparallel pairs than seen for RM734 and its nitrile terminated homologue. For DIO, the antiparallel form is the global energy minimum (Fig. 8a) albeit by only 0.1 kcal mol$^{-1}$, despite the fact polar order is known to prevail (as the material exhibits polar LC phases). For CIO and DIO(-F) the antiparallel form is far lower in energy than the parallel form. This highlights again that quite subtle differences in intermolecular interactions are responsible for the preference for parallel or antiparallel organization, which manifests in the presence (or absence) of polar liquid crystal phases.

Finally, we conclude our analysis by considering the bPES of two different molecules, RM734 and DIO. Chen *et al* showed that, experimentally, these display ideal mixing, despite the differences in their chemical structure, [59] and we have since suggested this may be a more general phenomenon. [43] Here, we find that the parallel configurations have larger complexation energies than the antiparallel form. For the mixed RM734/DIO system we find that the parallel form exhibits comparable intermolecular interactions to those seen for single component systems. In addition to π- π interactions and π-$NO_2$ interactions (Fig 9b), we find interactions between the nitro group and ester groups (Fig 9c) and between ester groups (Fig 9d-e). The ideal mixing of these materials reflects the fact the complementary nature of their intermolecular interactions. Framing the incidence of polar order in this way allows us to explain the rapid suppression of the $N_F$ phase in mixtures systems where this complementarity is lacking, for example 5CB and DIO/RM734. [60]

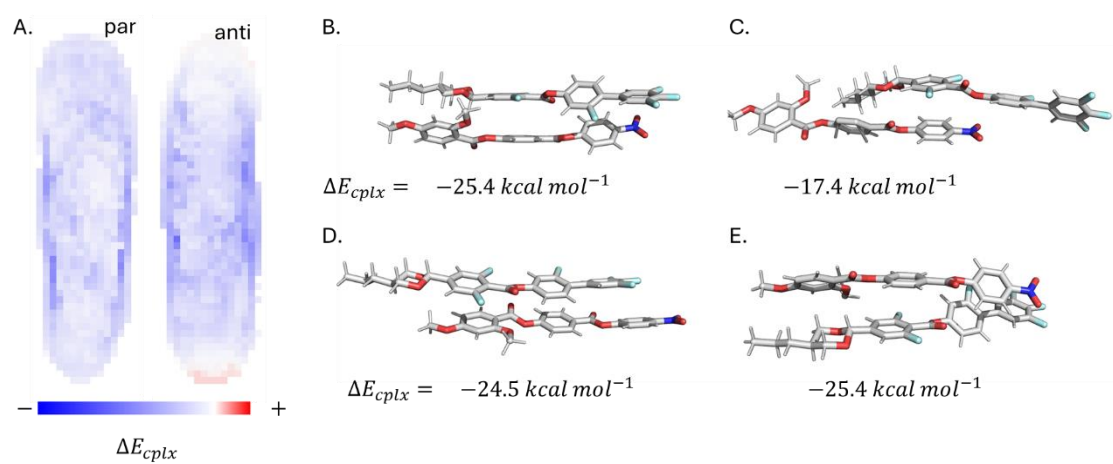

**Figure 9:** (A) Bimolecular potential energy surfaces for the RM734/DIO system in parallel and antiparallel configurations. Optimised minimum energy configurations of RM734/DIO (B-E). $\Delta E_{cplx}$ is the counterpoise complexation energy at the DFT B97D3/cc-pVTZ level.

**Conclusions**

We have developed a computational workflow for assessing how two molecules interact by first performing a rigid scan over the bimolecular potential energy surface and subsequently allowing relaxed geometry optimisation of a number of minima on this PES. We find that small changes in materials flip the energetic preference for (anti)parallel intermolecular orientations, for example swapping nitro (parallel) for cyano (antiparallel) in RM734-like materials. This allows us to explain the incidence/absence of polar order in terms of specific molecular interactions. In DIO-like materials, the fluorination pattern plays an important role in determining phase type through generation of anisotropies in the electrostatic potential of phenyl rings. Moreover, replacing dioxane with cyclohexane (aka CIO) is shown to reduce the complexation energy for parallel pairs, accounting for the lack of polar ordering in these materials. Our workflow is chemically specific; by describing the emergence of polar order in terms of interactions between molecules we have no need to invoke dipole-dipole forces, while steric factors are considered explicitly. Nevertheless, it may be premature to discount these effects entirely, as they doubtless influence the free energy of the system to some degree. Further, while we have not extended our analysis to all polar liquid crystals, the workflow here allows us to rationalise how specific structural changes favour (or disfavour!) (anti)parallel packing and thus polar order, and this offers the tantalising prospect of being able to understand counterintuitive effects resulting from fluorination patterns, as well as entropic effects.

**Acknowledgements**

R.J.M. thanks UKRI for funding via a Future Leaders Fellowship, grant number MR/W006391/1, and the University of Leeds for funding via a University Academic Fellowship. R.J.M. gratefully acknowledges support from Merck KGaA. Computational work was performed on ARC4 and AIRE, part of the high-performance computing facilities at the University of Leeds.